\documentclass{IOS-Book-Article}     
\usepackage{graphicx}
\usepackage{listings}
\usepackage[T1]{fontenc} 
\begin{document}
\begin{frontmatter}          
%
\title{Cudagrind: A Valgrind Extension for CUDA}
\runningtitle{Cudagrind}

%
\author[A]{\fnms{Thomas M.} \snm{Baumann}},
\author[A]{\fnms{Jos\'{e}} \snm{Gracia}}
\runningauthor{Thomas M. Baumann}
\address[A]{High Performance Computing Center Stuttgart, Nobelstr. 19, 70565 Stuttgart}
%

\begin{abstract}
Valgrind \cite{Valgrind}, and specifically the included tool Memcheck, offers an easy and reliable way for checking the correctness of memory operations in programs. This works in an unintrusive way where Valgrind translates the program into intermediate code and executes it on an emulated CPU. The heavy weight tool Memcheck uses this to keep a full shadow copy of the memory used by a program and tracking accesses to it. This allows the detection of memory leaks and checking the validity of accesses.

Though suited for a wide variety of programs, this approach still fails when accelerator based programming models are involved. The code running on these devices is separate from the code running on the host. Access to memory on the device and starting of kernels is being handled by an API provided by the driver being used. Hence Valgrind is unable to understand and instrument operations being run on the device.
 
To circumvent this limitation a new set of wrapper functions have been introduced. These wrap a subset of the CUDA Driver API function that is responsible for (de-)allocation memory regions on the device and the respective memory copy operations. This allows to check whether memory is fully allocated during a transfer and, through the functionality provided by Valgrind, whether the memory transfered to the device from the host is defined and addressable. Through this technique it is possible to detect a number of common programming mistakes, which are very difficult to debug by other means. The combination of these wrappers together with the Valgrind tool Memcheck is being called Cudagrind.
\end{abstract}


\end{frontmatter}


\section*{Introduction}

Checking of memory operations at runtime is an important tool in ensuring a programs correct behavior. Tools based on frameworks like Valgrind \cite{Valgrind} or Intel Pin \cite{IntelPin} and debuggers are valuable assets. A solid understanding of their correct usage can potentially save hundreds of hours spent on locating subtle problems like a misaligned pointer or prematurely freed memory regions. The problems arising from these type of errors include erratic and undefined behavior due to reading of undefined memory regions and crashes due to writing into forbidden memory areas. In the worst case this can lead to serious security concerns, for the user and the operating system alike, when the spurious access happens on memory that has been allocated by different parts of the system.

But despite the importance of these type of runtime checks the existing tools do not apply when working with accelerators like CUDA based GPGPUs from NVIDIA. In this case the kernels running on the device operate on dedicated device memory. This device memory features a very high bandwidth that allows exceptional performance boosts for the right problems. But at the same time it is very difficult to debug kernels running on the device, since the memory is not accessible from the host part of the program. In case of CUDA this memory can only be accessed through a certain set of functions offered by the CUDA driver and provided through it's driver API \footnote{http://docs.nvidia.com/cuda/cuda-driver-api/index.html} or from inside a kernel. Due to this indirect access pattern classical tools, like Valgrind, are unable to work with and on device memory. Tools working on the device, on the other hand, often do not handle the memory transactions between host and device.

This is where {\bf Cudagrind}, the tool presented in this paper, comes into play. It aims at establishing the missing link between host and device memory checking and is powered by the feature rich Valgrind framework.

\section{Memory Checking}

   
There's a wide variety of tools available when it comes to ensuring the correctness of a given program. These range from static analysis tools like Lint \cite{Lint}, over tools working at runtime like Valgrind \cite{Using_Valgrind}, to post-mortem analysis offered by debuggers which are invoked after a problem occurs. While these tools can't offer a full protection against memory errors of any kind, else they would solve the halting problem, they do offer an easy way to increase the confidence into ones own code. Especially with ever growing projects that feature hundreds of thousand lines of code which are impossible to check manually.

   \subsection{Valgrind: Heavy Weight Dynamic Binary Instrumentation}
   
   The Valgrind framework offers a novel way for dynamic instrumentation of arbitrary programs. When executing a program with Valgrind a just-in-time ({\bf JIT}) compiler translates its binary representation into an intermediate representation ({\bf IR}). This architecture-neutral IR, called VEX, is designed to be executed by a generic x86 processor which is provided and emulated by the Valgrind runtime. The JIT compiler ensures that each machine instruction in the original binary is translated into an equivalent sequence of VEX/IR instructions to be executed on the virtual processor. This approach comes with the downside of needing a working JIT compiler for each architecture Valgrind is being executed on. But it also allows dynamic instrumentation of any program at runtime.
   
   Tools build on top of the Valgrind framework, like the well-known memory checker Memcheck, utilize this dynamic instrumentation for various runtime checks. In the case of Memcheck the tool keeps a full shadow copy of the executed program's memory and registers. This copy is updated every time the program accesses memory regions or (de-)allocates memory. The information being tracked in this shadow copy allows Memcheck to perform checks on every access to any memory region or register. If there is an arbitrary access to unallocated memory or reading access to memory, that has been allocated but never defined, the tool can now print error messages and warnings at runtime. Due to it being a full copy the implemented checks are performed with bitwise precision.
   
   The downside of this approach lies in the sever overhead incurred at runtime. The emulation performed by the Valgrind runtime slows a program by a factor of 4 compared to its normal execution. Using a tool like Memcheck can raise this factor up to 100 due to the additional work performed by the tool. Additionally, depending on the type of instrumentation, the memory available to a program can be much smaller as well. For Memcheck the available memory is at least cut in half due to the full shadow copy being generated.

   \subsection{Cudagrind: The Missing Link}
   
   The way Valgrind handles the translation of machine code into the VEX IR poses another problem. The JIT compiler has to be adapted to every new architecture Valgrind is meant to work with. This is true for instructions introduced by new processor architectures, e.g. the AVX instructions introduced with Intel's Sandy Bridge architecture. But also accelerator based programing models that rely on dedicated devices, like NVDIA's CUDA GPGPUs or also Intel's Xeon Phi, are affected. 
   
   In these cases a precompiled binary, e.g. a CUBIN file, or dedicated intermediate code, e.g. PTX binary code, is being transfered to the accelerator. The execution of these binaries or kernels is then controlled through a set of functions offered by the underlying driver or runtime API. This prevents Valgrind from handling and instrumenting the code inside these kernels. The Valgrind JIT does not understand the CUBIN or PTX code and it does not have access to code running solely on the device.
   
   This gap is meant to be closed by the memory checker \verb|cuda-memcheck|, provided with NVIDIA's CUDA SDK. But despite working well with the actual code running on the device, there's still a missing link. \verb|Cuda-memcheck| is unable to check the memory transfers happening between host and device, as these are handled through a set of functions accessible only through the CUDA Driver API. Valgrind on the other hand is unable to check the correctness of these operations since the Valgrind runtime does not know about the memory located on the card and whether it's been previously allocated or not.
   
   This is where Cudagrind comes into play. A set of wrappers, based on Valgrind's wrapping functionality, is provided. It covers all CUDA Driver API functions related to allocation, deallocation and transfer of memory between host and device. The wrappers perform internal bookkeeping of allocated memory on the device, check the CUDA Driver API function parameters at runtime. Additionally definedness checks, provided by Valgrind, are utilized in order to locate errors happening during memory transfers between host and CUDA devices.

\section{Cudagrind Internals} 

\begin{figure}[ht]
   \includegraphics[width=.75\textwidth]{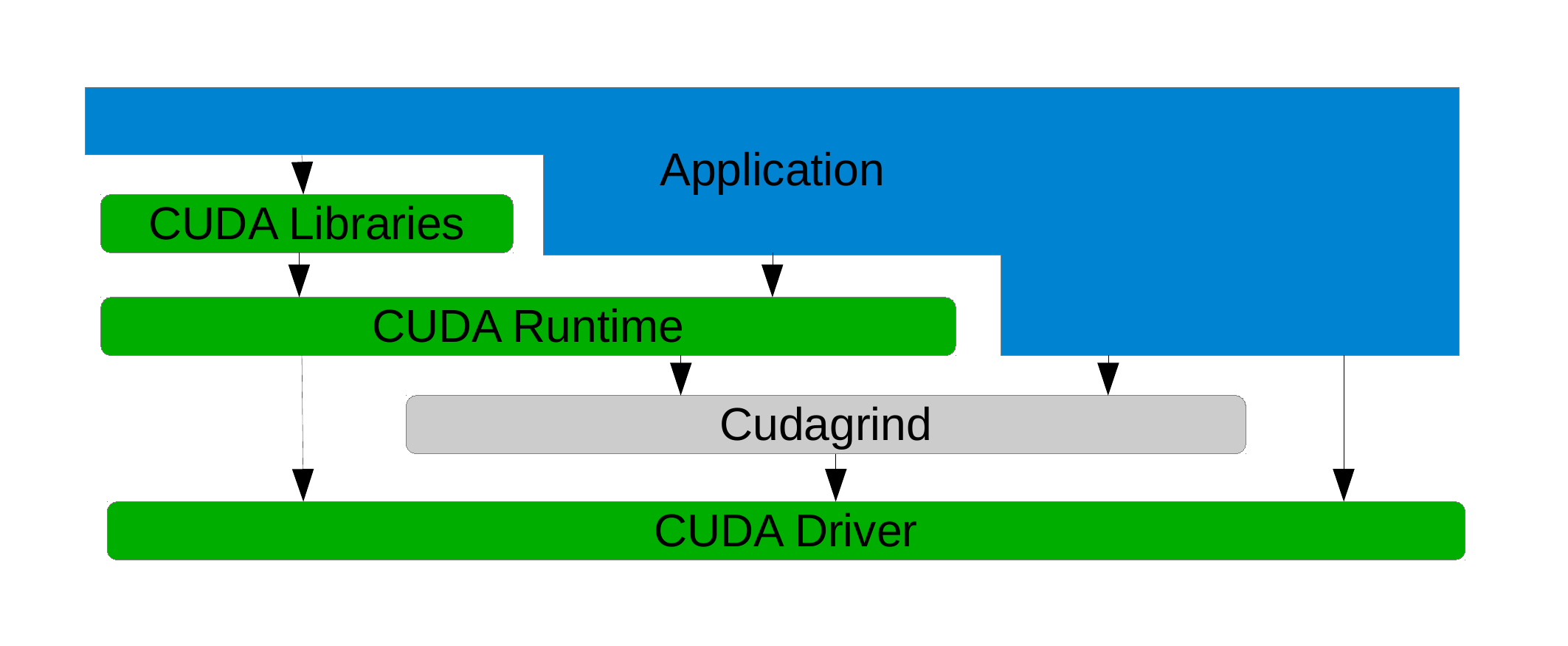}
   \caption{The placement of Cudagrind in the CUDA stack. There is a wrapper for every memory related function on the CUDA driver level. Functions calls from an application on a higher level, e.g. a CUDA library function, are handled when the respective driver function is called from the library or runtime level.}
   \label{fig:CUDA-Stack}
\end{figure}

In case of the CUDA programming model any access to a CUDA enabled device is handled by the CUDA driver. This is especially true for programs written fully on the driver level, but also holds when working on a higher layer in the CUDA stack (Figure \ref{fig:CUDA-Stack}). The CUDA runtime, which offers a more streamlined and easier to use access to the device, relies directly on the functions provided by the CUDA driver. So a call to \verb|cudaMemcpy(..)| implicitly leads to a call of any of the \verb|cuMemcpyXtoY(..)| functions provided by the driver. Moving one abstraction layer higher offers the same picture. Any program utilizing a pragma based approach like OpenACC will rely on the functionality offered by the CUDA driver or, implementation dependent, also the CUDA runtime. Finally any program relying on a CUDA library will implicitly rely on the underlying implementation, which in turn relies on the CUDA driver to provide the needed functionality.

\begin{figure}[ht]
   \includegraphics[width=.75\textwidth]{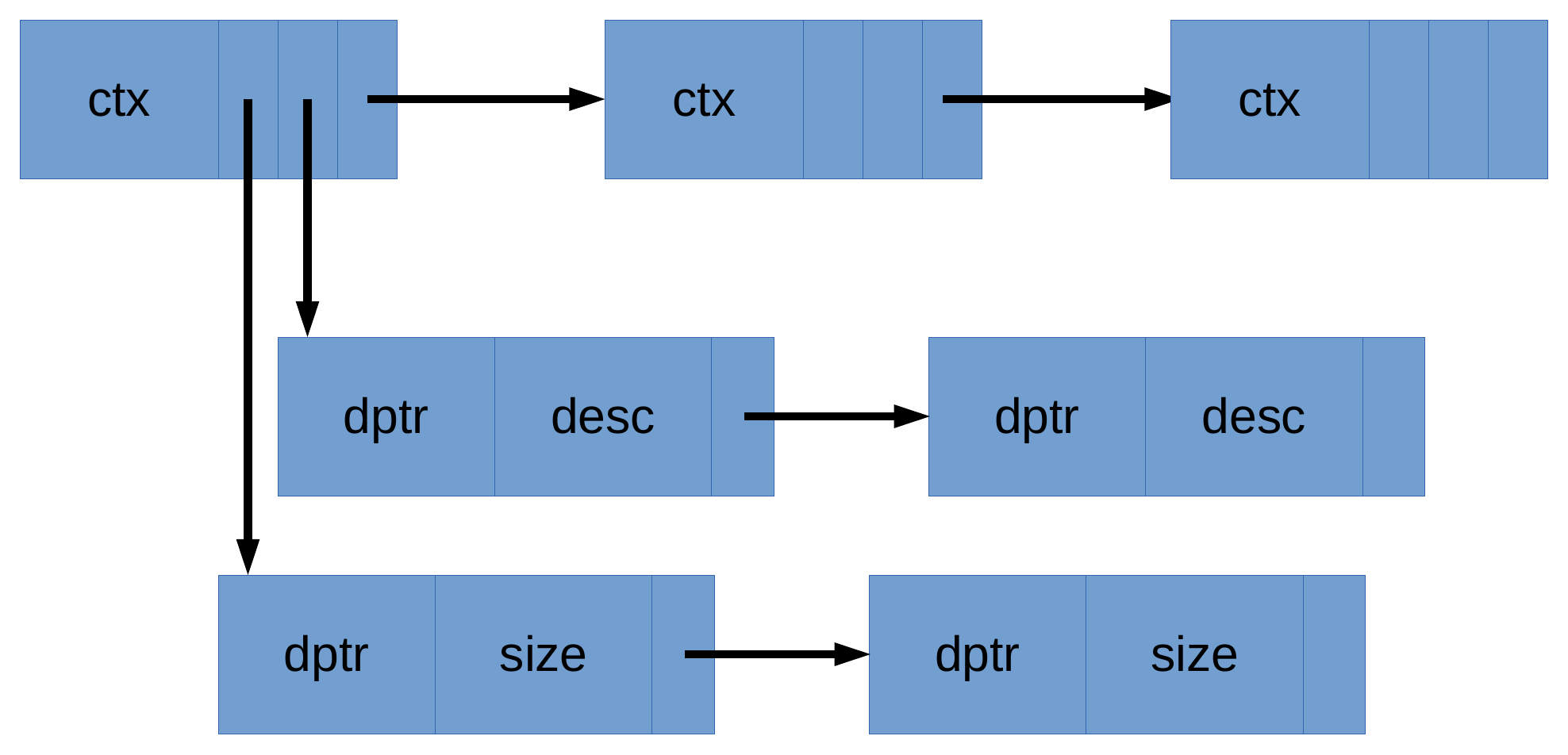}
   \caption{Internal list of memory regions that have been allocated on the device. Each allocation generates a new entry into the list of linear memory or device arrays of the respective CUDA context ctx. Both lists contain information about the device pointer returned by the allocation and either the size of the linear memory or device array descriptor.}
   \label{fig:CG-List}
\end{figure}

So no matter the programing model or abstraction layer used, every program relies on the CUDA driver. Since this is also true for any device memory related operation, the CUDA driver is a natural choice for the wrappers provided by Cudagrind. Anytime a program handles memory on the device or transfers data between the device and/or the host the respective Cudagrind wrapper will be called instead. Depending on the requested operation the internal list of memory regions allocated on the device is being updated (Figure \ref{fig:CG-List}) and the function parameters are cross checked against this list as well as the knowledge of Valgrind about the device on the host memory. This allows detection of various errors like

\begin{itemize}
   \item data being copied into or from unallocated memory regions.
   \item undefined data being copied from host onto the device\footnote{This might not be an error in cases where the host memory is accessed in a strided way.}.
   \item copying more data than there's allocated memory.
   \item certain concurrent accesses when several threads are used.
\end{itemize}

\section{Software Requirements}\label{sec:req}

To compile the Cudagrind wrappers the following components need to be installed and available

\begin{itemize}
   \item Valgrind 3.6.0 or newer.
   \item CUDA SDK 4.0 or newer.
   \item CUDA driver that works with installed SDK.
\end{itemize}

Additionally the location of the respective driver's \verb|libcuda.so| and Valgrind's \verb|valgrind.h| needs to be known and passed to the linker when compiling Cudagrind.

To execute a program with the Cudagrind wrappers the same Valgrind version used during the compilation needs to be present. The generated \verb|libcudagrind.so| and the \verb|libcuda.so| from the respective CUDA driver needs to be added to \verb|LD_PRELOAD| in this order. The addition of \verb|libcuda.so| to \verb|LD_PRELOAD| is only needed in cases where the original program has not been fully linked against it. This is the case when using certain OpenACC compilers, where running the program with the Cudagrind wrappers would lead to unknown references to functions called only from within the wrappers but not the main program itself.

Now the executable that is to be run with Cudagrind can simply be called with \verb|valgrind <executable>|. Every time one of the wrapped functions in the CUDA driver is executed Valgrind will replace it with the respective Cudagrind wrapper and perform all needed checks at runtime. Optionally a suppression file can be generated and passed to Valgrind at runtime. This suppresses certain spurious errors reported by Valgrind in the CUDA driver\footnote{See the Valgrind manual that comes with your installation or at the Valgrind homepage at http://valgrind.org/ for more information about suppression files.}. In certain rare cases the lib path of the Valgrind installation \verb|/path/to/valgrind/lib/valgrind| needs to be added to \verb|LD_LIBRARY_PATH|.

\section{Usage Example}

This section presents a brief introduction to how Cudagrind can assist in detecting and solving memory related errors in a CUDA program. Let 'vecsum' in Listing \ref{lst:vecsum} be a kernel that performs a basic vector addition of two input vectors \verb|a| and \verb|b| and writes the result into the vector \verb|c|.

\begin{figure}[ht]
  \lstset{basicstyle=\scriptsize,language=C,frame=single,caption=Basic vector sum in CUDA C.,label=lst:vecsum}
  \begin{lstlisting}
__global__ void vecsum(double *c, double *a, double *b) {
   start = threadIdx.x+(blockIdx.x*blockSize.x);
   size  = blockSize.x*grindSize.x;
   for (int i = start ; i < size ; i += blockSize.x) {
      c[i] = a[i]+b[i];
   }
}
  \end{lstlisting}
\end{figure}

Further assume the following excerpt from a program utilizing this kernel in Listing \ref{lst:vecsum}, where a wrong amount of memory is allocated for the result pointed at by \verb|c|. In this case the programmer only allocated enough memory to hold \verb|size| values of the type float instead of the twice as big type double expected by the kernel.

\begin{figure}[ht]
   \lstset{basicstyle=\scriptsize,language=C,frame=single,caption=Part of a CUDA C program utilizing the vecsum kernel from Listing \ref{lst:vecsum}. Note the faulty allocation of the result variable c using {\bf float} instead of {\bf double}.,label=lst:snippet}
   \begin{lstlisting}
...
double *c, *a, *b;
double *c_host;
int size = 1000000;
...
cudaMalloc((void**)&c, size*sizeof(float));
cudaMalloc((void**)&a, size*sizeof(double));
cudaMalloc((void**)&b, size*sizeof(double));
...
vecsum<<<gridSize, blockSize>>>(c, a, b);
...
cudaMemcpy(c_host, c, size*sizeof(double), 
           cudaMemcpyDeviceToHost);
...
   \end{lstlisting}
\end{figure}

The result of executing this code with different types of error checking is now examined in the following sections.

   \subsection{No Tools, No API Error Value Checking}
   
   If the code is being run, without external tools or utilization of error values returned by the CUDA runtime, this error might go unnoticed. First the kernel's result will be undefined due to the write into unallocated memory regions. Then the call to \verb|cudaMemcpy| will fail due to the wrong amount of memory being scheduled for copying. In the best case this will lead to a failure detected in the remainder of the program. But more likely the bug will go unnoticed, leading to the program continuing with undefined values stored in the host memory pointed at by \verb|c_host|. On the bright side, this method of error detection does not ask for recompilation nor debug information being present in the binary!

   \subsection{No Tools, API Error Value Checking}
   
   The simplest, yet still often not utilized, method of error checking within any program relying on the CUDA runtime is the error value returned by the runtime. This can be retrieved either directly as return value from each function call or, especially in case of asynchronous operations, by a call to \verb|cudaGetLastError| or \verb|cudaPeekAtLastError|. For asynchronous operation a preprocessor macro like the one in Listing \ref{lst:API-check} can be used to simplify the process. To get the same functionality for the asynchronous case a synchronization would have to be added to get the exact error message generated by the operation.
   
   \begin{figure}[ht]
      \lstset{basicstyle=\scriptsize,language=C,frame=single,caption=Preprocessor macro to simplify runtime API error value checking. Use cudaVerify with every call to the CUDA runtime. Similar principle applies to CUDA driver API error checking. This and similar macros are included in the cutil.h header file of CUDA SDKs prior to 4.2.,label=lst:API-check}
      \begin{lstlisting}
#define cudaVerify(x) do {                        \
 cudaError_t __cu_result = x;                     \
 if (__cu_result!=cudaSuccess) {                  \
  fprintf(stderr,                                 \
     "%s:%i: error: cuda function call failed:\n" \
     "  %s;\nmessage: %s\n",__FILE__,__LINE__,    \
     #x,cudaGetErrorString(__cu_result));         \
  exit(1);                                        \
 }                                                \
} while(0)
      \end{lstlisting}
   \end{figure}
   
   Using \verb|cudaVerify| on the call to \verb|cudaMemcpy| in Listing \ref{lst:snippet} produces the output shown in Listing \ref{lst:API-check_out}. This method shows the location in the code where the error is happening, even if the binary contains no debug information, and causes nearly no overhead. But the error message 'invalid argument' reported by the CUDA runtime is very undescriptive and the method is error prone since the \verb|cudaVerify| macro has to be used with every call to any CUDA runtime function. Additional care has to be taken for asynchronous functions and when the code also utilizes the CUDA driver API or other CUDA based programming models.
   
   \begin{figure}[ht]
      \lstset{basicstyle=\scriptsize,frame=single,caption=Running the faulty code shown in Listing \ref{lst:snippet} with verbose runtime error checking produces the following output. Location of the error is shown{,} but the error message is non-descriptive.,label=lst:API-check_out}
      \begin{lstlisting}
hpc43598 n093302 306$./a.out
example01.cu:60: error: cuda function call failed:
   cudaMemcpy(c_host, c,(vector_size)*sizeof(double),
cudaMemcpyDeviceToHost);
message: invalid argument
      \end{lstlisting}
   \end{figure}

   \subsection{Valgrind/Memcheck}
   
   Utilizing only the Valgrind tool Memcheck produces no warnings or error messages, except possibly spurious errors inside the CUDA driver library \verb|libcuda.so|. As mentioned in Section \ref{sec:req} these are not connected to actual errors in the program being run and can be either ignored or suppressed with a Valgrind suppression that's tailored to CUDA driver being used. In this case Valgrind is unable to detect the error since the problem only occurs on the device and inside the kernel being executed. This part of the program is effectively invisible to the Valgrind framework. At the current time it is unable to translate and run the PTX or CUBIN code that is being transfered to and run on the device by the CUDA driver.

   \subsection{cuda-memcheck}
   
   NVIDIA's memory checker \verb|cuda-memcheck| can also be used to check for memory related problems at runtime. In this case the program is passed to \verb|cuda-memcheck| as parameter and the tool prints every error located to \verb|stdout|. Another plus point is the tight integration with \verb|cuda-gdb|. Using '\verb|set cuda memcheck on|' inside the \verb|gdb| based debugger, which also comes bundled with the CUDA SDK, will enable the same checking abilities when debugging a program.
   
   Listing \ref{lst:cuda-memcheck_out} shows the output produced by \verb|cuda-memcheck| when being used on the code from Listing \ref{lst:snippet}. In this case \verb|cuda-memcheck| manages to detect location of the error and reports the internal error number reported by the CUDA runtime. But the display of the actual error name or description is missing as is the actual location in the source code of the program. The location is only reported as address in the binary, not in the source code, even if the program has been compiled with \verb|-g -G| for full debug informations.
   
   \begin{figure}[ht]
      \lstset{basicstyle=\scriptsize,frame=single,caption=Running the faulty code shown in Listing \ref{lst:snippet} with cuda-memcheck produces this output. Location of the error is only shown inside the binary and only the error number is reported.,label=lst:cuda-memcheck_out}
      \begin{lstlisting}
hpc43598 n093302 307$cuda-memcheck ./a.out
CUDA-MEMCHECK
Program hit error 11 on CUDA API call to cudaMemcpy
Saved host backtrace up to driver entry point at error
Host Frame:/../libcuda.so [0x26a180]
Host Frame:/../libcudart.so.5.0 (cudaMemcpy + 0x28c) [0x3305c]
Host Frame:./a.out [0xc55]
Host Frame:/lib64/libc.so.6 (__libc_start_main + 0xfd) [0x1ecdd]
Host Frame:./a.out [0x9f9]
ERROR SUMMARY: 1 error
      \end{lstlisting}
   \end{figure}

   \subsection{Valgrind/Memcheck with Cudagrind Wrappers}
   
   Running the aforementioned code snippet in Valgrind with enabled Cudagrind wrappers will produce the output shown in Listing \ref{lst:Cudagrind_out}. The output shown gives a detailed error message. It tells the user that the allocated device memory during a device to host transfer has been too small and that there are only half of the needed amount of bytes available. This already gives a hint at the actual error in Listing \ref{lst:snippet}, where the result vector \verb|c| has been allocated for floats instead of doubles. Additionally, if the program has been compiled with debug information, the exact location in the source code is being shown as in Listing \ref{lst:Cudagrind_out}. Otherwise, similar to running the program with \verb|cuda-memcheck|, the location inside the binary is being shown.
   
   \begin{figure}[ht]
      \lstset{basicstyle=\scriptsize,frame=single,caption=Running the faulty code shown in Listing \ref{lst:snippet} with the Cudagrind wrappers produces this output. Location of the error in the source code is only shown when compiled with debug informations. A detailed error message gives a hint about the potential problem.,label=lst:Cudagrind_out}
      \begin{lstlisting}
hpc43598 n093302 307$valgrind ./a.out
Error: Allocated device memory too small for device->host copy.
Expected 8000000 allocated bytes but only found 4000000.
 at 0x4C18E79: VALGRIND_PRINTF_BACKTRACE (valgrind.h:4477)
 by 0x4C19261: cuMemcpyDtoH_v2 (cuMemcpyDtoH.c:58)
 by 0x5A1E531: ??? (in /../libcudart.so.5.0.35)
 by 0x5A40E43: cudaMemcpy (in /../libcudart.so.5.0.35)
 by 0x400C54: main (example01.cu:60)
      \end{lstlisting}
   \end{figure}

\section{Summary}

Cudagrind offers a novel approach to close the missing link between host side memory checking, e.g. through Valgrind based tools, and device side memory checking with \verb|cuda-memcheck|. Utilizing the provided set of wrappers it is now possible to track the allocation status and definedness of memory during host/device, device/host and device/device memory transactions at runtime. With Cudagrind being dynamically executed by the Valgrind core and depending on only the CUDA driver there is no recompilation needed, except for the precise localization of potential errors in the original source code. 

Due to the location of the Cudagrind wrappers in the CUDA stack they work with 'everything CUDA'. This includes programs directly accessing the CUDA driver or the CUDA runtime as well as applications relying on CUDA libraries or based on pragma based programming models like OpenACC. The only requirement is for the chosen method to be built on top of the CUDA driver. Additionally with how Valgrind and its wrappers work there is no overhead when running the program on its own, as the original program does not need to be changed or recompiled/-linked in any way.

At the time of writing a beta Version of Cudagrind has been released\footnote{https://www.hlrs.de/organization/av/spmt/research/cudagrind/}.
  
  \subsection{Outlook}
  
  With the work presented in this paper a solid foundation for memory transaction checking for CUDA based programs has been laid out. Future version might provide additional wrappers that, in addition to the actual memory transactions, check the arguments of kernels running on the device. This involves providing a wrapper for \verb|cuLaunchKernel| and unpacking the parameter list passed to this function in order to link these parameters to the actual variables in the program.
  
  Another possible approach would be a thorough integration of CUDA within the Valgrind framework. Akin to how Valgrind provides a virtual x86 CPU it would be possible to extend it's core with a virtual device running PTX or CUBIN code produced by NVIDIA's CUDA compiler. While allowing full control and complete memory checking capabilities, even for code running on the device, this is no trivial task though. One way to achieve this goal might lie in the extension and integration of a dynamic CUDA compilation framework like 'GPU Ocelot' \cite{Ocelot}. This would compile CUDA code into host based code, which in turn could be handled by a modified Valgrind kernel which would have to be aware of the properties specific to code running on a device. Another approach, as proposed in \cite{Ocelot_Instrumentation}, would be to include the instrumentation in Ocelot. But in this case a solution for memory checking on the host side would've to be added to Ocelot.

\acknowledgements{This work was supported  by the H4H
project funded by the German Federal Ministry for Education and Research
(grant number 01IS10036B) within the ITEA2 framework (grant number 09011).}


  %

\bibliographystyle{unsrt} 
\bibliography{ParCo-2013}


\end{document}